# An atomistic quantum transport solver with dephasing for field-effect transistors

HASSAN RAZA and EDWIN C. KAN

*School of Electrical and Computer Engineering, Cornell University, Ithaca, NY 14853 USA*

hr89@cornell.edu

**Abstract.** Extended Hückel theory (EHT) along with NEGF (Non-equilibrium Green's function formalism) has been used for modeling coherent transport through molecules. Incorporating dephasing has been proposed to theoretically reproduce experimental characteristics for such devices. These elastic and inelastic dephasing effects are expected to be important in quantum devices with the feature size around 10nm, and hence an efficient and versatile solver is needed. This model should have flexibility to be applied to a wide range of nano-scale devices, along with 3D electrostatics, for arbitrary shaped contacts and surface roughness. We report one such EHT-NEGF solver with dephasing by self-consistent Born approximation (SCBA). 3D electrostatics is included using a finite-element scheme. The model is applied to a single wall carbon nanotube (CNT) cross-bar structure with a $C_{60}$ molecule as the active channel. Without dephasing, a negative differential resistance (NDR) peak appears when the $C_{60}$ lowest unoccupied molecular orbital level crosses a van Hove singularity in the 1D density of states of the metallic CNTs acting as contacts. This NDR diminishes with increasing dephasing in the channel as expected.

## 1. Introduction

Interest in atomic-scale conduction is driven by both scientific curiosity and technology advancement. At the scale below 20nm, atomistic transport models with flexible handling of dephasing mechanisms at room temperature will be very useful. We report here an EHT-NEGF model (extended Hückel theory, Non-equilibrium Green's function) which can be used to calculate the transport through ultra-small devices with realistic contact composition [*e.g.* carbon nanotube (CNT), Si(001), Au(111) and STM tip] in the presence of elastic and inelastic dephasing, which is incorporated by self-consistent Born approximation (SCBA). This model consists of an atomistic channel, having dimensions on the order of few nano-meters, with source/drain contacts and up to four arbitrary shaped gates as shown in Fig. 1. In this article, we use the CNT cross-bar structure for illustration as shown in Fig. 2. Carbon nanotube contacted molecules have been a topic of recent study [1].

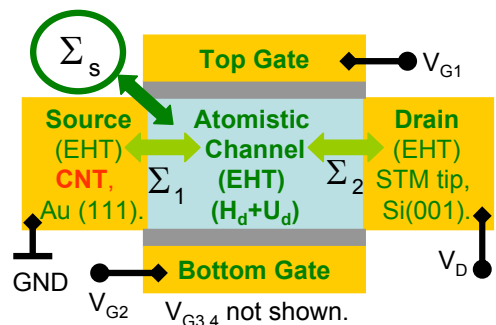

*Figure 1.* The generic device structure consists of a channel contacted by source and drain, which can be either CNT, STM tip, Au(111) or Si(001). The gates can be of any shape and material. $U_d$ consists of 3D Laplace potential due to the applied drain/gate voltages and Poisson potential due to the non-equilibrium carrier statistics.

## 2. EHT-NEGF-SCBA Formalism

The electronic structure of choice for the solver is EHT, which is a non-orthogonal tight-binding (TB) scheme with a well-defined double zeta Slater-Type Orbital (STO) basis set and has been used with success for nano-scale structures [2,3,4,5,6]. For quantum transport, we use NEGF formalism. Our model is derived from the previous Hückel-IV 3.0 [6], which is an EHT-NEGF solver with Au(111) contacts. There are two new enhancements: (1) more contact compositions, e.g. CNT [2], Si(001) [3,4] and STM tip [4,5]; (2) inclusion of elastic and inelastic dephasing within SCBA [4,5,7] which captures electron-phonon scattering.

We define the time-retarded Green's function as:

$$G = \left[(E + i0^+)S - H_d - U_d - \Sigma_c - \Sigma_s\right]^{-1} \quad (1)$$

where $H_d$ is the bare device Hamiltonian and $S$ is the overlap matrix due to non-orthogonal basis set. $U_d$ incorporates both 3D Laplace potential (which acts as an electrostatic boundary condition) due to applied drain and gate voltages and Hartree potential due to non-equilibrium carrier statistics in the device region. $\Sigma_c(c=1,2)$ is the contact self-energy which defines electronic boundary conditions for the Hamiltonian and $\Sigma_s$ is the scattering self-energy which defines scattering boundary conditions and is given as [4,5]:

$$\Sigma_s(E) = \frac{1}{\pi}\int_{-\infty}^{\infty} dy \frac{-\Gamma_s(y)/2}{E - y} - i\frac{\Gamma_s(E)}{2} \quad (2)$$

where $\Gamma_s$ is the scattering broadening function. The real part of $\Sigma_s$ involves a Hilbert transport of $\Gamma_s/2$ and is computationally very expensive due to integration over a wide energy range. If the density of states (DOS) of the contact is uniform over this range, the real part is usually negligible. This may not be the case for CNT due to singularities in DOS. However, for numerical convenience we ignore it here. $\Gamma_s$ is defined as [4,5,7,8]:

$$\Gamma_s = \int_0^{\infty} \frac{d(\hbar\omega)}{2\pi} D^{em} S[G^n(E + \hbar\omega) + G^p(E - \hbar\omega)]S$$
$$+ D^{ab} S[G^n(E - \hbar\omega) + G^p(E + \hbar\omega)]S \quad (3)$$

where $D^{em}$ and $D^{ab}$ are emission and absorption dephasing functions defined as $D^{em}=(N+1)D_o$ and $D^{ab}=ND_o$ and are related by $D^{ab}=D^{em}e^{-\hbar\omega/kT}$, given $N$ is Bose factor from the equilibrium phonon occupation for the phonon mode of $\hbar\omega$ energy. The general procedure for calculating $D_o$ is outlined in [7]. For elastic dephasing, Eq. 3 reduces to [4,5]:

$$\Gamma_s = \int_0^{\infty} \frac{d(\hbar\omega)}{2\pi}\left[D^{em} + D^{ab}\right] \cdot SAS = D \cdot SAS \quad (4)$$

where $A=G^n+G^p$ is the spectral function and D is dephasing strength − a fourth ranked tensor but phenomenologically approximated by a scalar in this work. Approximating D by a scalar treats $T_1$ and $T_2$ times in an average manner and seems to reproduce experiments [4]. $G^n$ and $G^p$ are the electron and hole correlation functions, respectively. $G^n$ relates to density matrix as:

$$\rho^n = \frac{1}{2\pi}\int_{-\infty}^{\infty} dE G^n(E) \quad (5)$$

Since $\Sigma_s$, $G$, $G^n$, and $\Sigma_s^{in,out}$ [4,5,7] depend on each other, one needs to solve for these in a self-consistent manner giving this scheme the name SCBA.

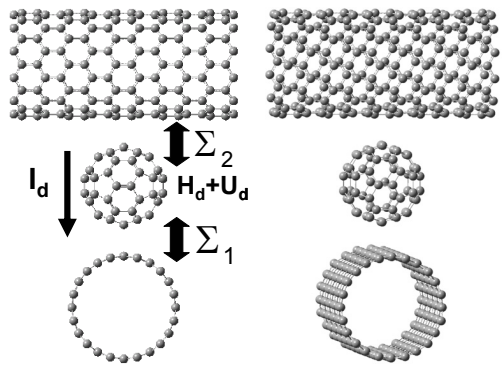

Figure 2. The ball and stick model of the cross-bar device. A $C_{60}$ is the active channel contacted by two (12,0) zigzag metallic CNTs which function as source and drain. The $C_{60}$ to CNT spacing is taken to be about 3Å, which is close to the C-C $\pi$ stacking distance. The current flows from the source CNT to the drain CNT though the $C_{60}$. An alternate view is provided of the same structure for clarity [12].

The contact self-energy $\Sigma_c$ is given as $[(E+i0^+)S_{dc}-H_{dc}]g_{s-c}[(E+i0^+)S_{cd}-H_{cd}]$, where $g_{s-c}$ is the contact's surface Green's function. For the single-wall CNT, Green's function of the CNT is itself the surface Green's function. We perform this calculation and the resulting DOS of the (12,0) zigzag CNT is shown in Fig. 3. As expected there are van Hove singularities in the 1D DOS which play an important role in the transport results. One important observation is the curvature induced band gap in otherwise metallic CNT. Although small, but it depicts the advantage of EHT over other simpler TB methods where such effect cannot be reproduced. Furthermore, the location of the three-fold LUMO level of $C_{60}$ is set to about 0.55eV above $\mu_o$ for demonstration. The LDA LUMO level for $C_{60}$ in the gas phase is close to zero energy. Bonding with CNT is expected to cause some charge transfer and hence shifts the LUMO level to higher energies. A different LUMO level position would result in different voltage at which NDR occurs.

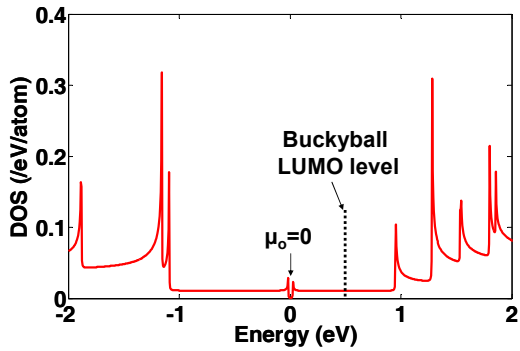

*Figure 3. DOS of (12,0) zigzag CNT showing van Hove singularities. A curvature induced band gap is observed which cannot be obtained by simpler TB calculations. The three-fold LUMO level of $C_{60}$ is about 0.55eV above $\mu_o$ for demonstration.*

## 3. 3D Electrostatics

The effect of applied drain/gate voltage is included in the EHT Hamiltonian in the form of a 3D Laplace potential, which is calculated by the finite element solver in COMSOL for contacts with arbitrary shapes. This feature is important because at nano-scale, not only electrostatic effects of surface roughness may become important, but also contacts may be intentionally engineered to have certain shape. One example of such a contact is CNT, where the contact surface has a curvature. A 2D voltage profile of the 3D Laplace solution for the CNT cross-bar structure is shown in Fig. 4. Clearly, there is curvature induced non-linear drop between CNTs. This non-uniformity in the voltage drop may become more important as we scale down the diameter of the contacts.

The Hartree potential is solved using complete neglect of different overlap (CNDO) [4,5,6]. This method is physically rigorous and incorporates exchange and correlation effects. Since density matrix changes with applied bias and *vice versa*, we have to solve for the two quantities self-consistently. Once self-consistency is achieved, the current through a contact is:

$$I_c = \frac{2e}{h} \int_{-\infty}^{\infty} dE tr [\Sigma_c^{in} A - \Gamma_c G^n] \qquad (6)$$

For SCBA, the current through the scattering 'contact' is always zero. It has been previously reported, that the Hartree potential shifts the energy levels [4,5], which will shift the voltage at which NDR occurs and will broaden the NDR peak due to the charging effect.

## 4. Results

For the cross bar structure as shown in Fig. 2, the resulting $I_d$-$V_d$ at 300K is shown in Fig. 5. There is a clear NDR peak observed with peak to valley ratio of about 10. This feature is similar to the ones reported in [9,10]. The physics behind this NDR phenomenon is explained in Fig. 6. For low $V_d$ (*e.g.* -0.9V), the three-fold degenerate LUMO level is off-resonant with the van Hove singularity and hence transmission is small. At about -1V, the level is in resonance resulting in large transmission due to strong coupling to the contact as shown in Fig. 6(a). With increasing voltage, the level slips past the singularity and hence transmission becomes small resulting in decrease in current and hence NDR.

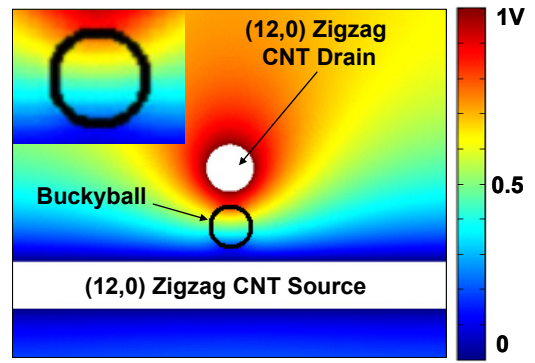

*Figure 4. 3D electrostatics of the cross-bar structure. The metallic CNTs are modeled as metallic cylinders. The $C_{60}$ is modeled as a spherical dielectric having $\varepsilon_r$=2 and is surrounded by a high-K dielectric having $\varepsilon_r$=12. A 2D voltage profile though the middle of the $C_{60}$ is shown. There is highly non-linear voltage drop between the two CNTs in the device region as shown more clearly in the inset.*

We further show in Fig. 5 that elastic dephasing smears out these NDR features. Figures 6(b,c) show the corresponding transmission plots. The difference in transmission between -1V and -1.1V is not large, which results in a smeared resonance. The physical reason behind this is that dephasing broadens the spectral density of the LUMO level as shown in Fig. 6(d). Since the molecular level is broadening over an appreciable energy range, the resonance with the singularity is not peaked at a particular energy. This elastic dephasing could be due to the low-energy acoustic phonon modes resulting from the center-of-mass motion of the $C_{60}$ with respect of the CNT contacts. It has been reported previously [11] that this phonon mode has energy of about 5meV for a $C_{60}$ placed between gold contacts. It is not clear what this frequency would be for CNT contacts, but it is still expected to be small due to high mass of $C_{60}$ and given that $\omega \propto 1/\sqrt{m}$. If the energy of this mode is much less than 25meV at 300K, its contribution to transport can be approximated by elastic dephasing, because inelastic dephasing would give the same IV characteristics in the system we consider, however with a greater computational complexity. Contribution of any optical modes due to C-C bonds should be included using inelastic scattering and can be readily incorporated in our model.

inelastic dephasing. We have applied it to model a CNT cross-bar structure with $C_{60}$ as the channel and have reported that NDR would occur when LUMO level crosses a singularity in CNT DOS. The voltage at which it happens depends on the relative gap between the LUMO level and the singularity. Furthermore, elastic dephasing due to acoustic phonon modes is expected to smear out these NDR features.

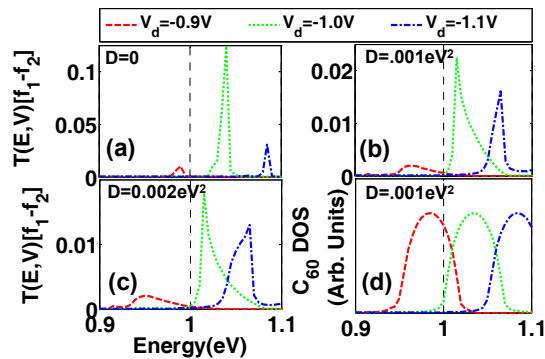

*Figure 6. Physical explanation of NDR. (a) Transmission through three-fold degenerate LUMO level is small when away from the van Hove singularity at $V_d$=-0.9V. For $V_d$=-1V, the LUMO level is aligned with the van Hove singularity resulting in increased transmission due to stronger coupling with source contact. At $V_d$=-1.1V, the LUMO level is no longer aligned and hence transmission decreases resulting in a NDR peak. (b,c) Increasing dephasing results in wider transmission peaks due to the broadened DOS. (d) Corresponding DOS plots of $C_{60}$ LUMO level showing spectral broadening due to dephasing. Also, note that DOS shifts by half of $V_d$.*

We thank D. Kienle for useful discussions about CNT modeling. It is a pleasure to acknowledge F. Zahid and T. Raza for Hückel-IV 3.0 codes and later for visualization. The work is supported by National Science Foundation (NSF) and Nanoelectronics Research Institute (NRI) through Center for Nanoscale Systems (CNS) at Cornell University.

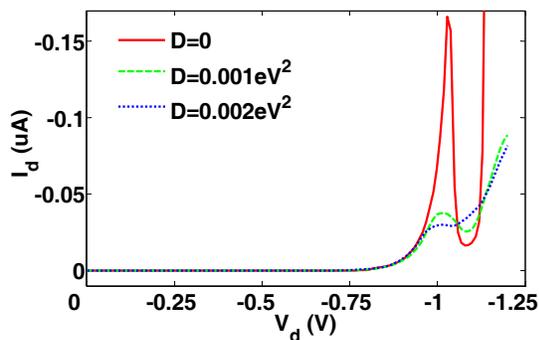

*Figure 5. The effect of elastic dephasing on $I_d$-$V_d$ characteristics of device in Fig. 2 at 300K. A negative differential resistance (NDR) event is due to the resonant tunneling when the three-fold degenerate LUMO level crosses the van Hove singularities in the 1D DOS of the CNT contacts. The observed peak-to-valley ratio is about 10. Elastic dephasing in the device region broadens the spectral density of $C_{60}$ and hence results in smearing out of the NDR.*

## 5. Conclusions

We have reported a versatile and computationally modest atomistic transport solver with elastic and